\documentclass[12pt]{article}

\usepackage[utf8]{inputenc}
\usepackage{latexsym, graphicx} 
\usepackage{amsmath,amsfonts,amssymb}
\usepackage{slashed}
\usepackage{cancel}
\usepackage{mathrsfs}
\usepackage{tensor}
\usepackage{xcolor}
\usepackage{cite}
\usepackage[colorlinks=true,linkcolor=blue,citecolor=blue,urlcolor=blue,filecolor=black]{hyperref}

\renewenvironment{thebibliography}[1]{
	\begin{oldthebibliography}{#1}
		\setlength{\itemsep}{2pt}
		\setlength{\parskip}{2pt}
	}
	{
	\end{oldthebibliography}
}

\numberwithin{equation}{section} 

\newcommand*{\dd}{\mathop{}\!d}

\begin{document}
	
	\begin{titlepage}
		\thispagestyle{empty}
		
		\begin{flushright}
		\end{flushright}
		
		\vskip1cm
		
		\begin{center}  
			{\Large\textbf{Hilbert space in the flat limit of AdS/CFT\\
					\vskip 2mm I.~Massive particles}}
			
			\vskip1cm
			
			\centerline{Kevin Nguyen}
			
			\vskip1cm
			
			{\it{Universit\'e Libre de Bruxelles and International Solvay Institutes,\\ ULB-Campus Plaine
					CP231, 1050 Brussels, Belgium}}\\
			\vskip 1cm
			{kevin.nguyen2@ulb.be}
			
		\end{center}
		
		\vskip1cm
		
		\begin{abstract} 
			The flat limit of AdS$_4$ spacetime corresponds, at the level of the symmetry algebra, to the Inönü--Wigner contraction $\mathfrak{so}(2,3) \to \mathfrak{iso}(1,3)$. We describe the emergence of the Hilbert space of massive scattering states of arbitrary integer spin $s$ from unitary lowest-weight representations $\mathcal{D}(\Delta,s)$ of the conformal group, using a basis of `pseudo-momentum' states first introduced by Fronsdal. These states nicely reduce to Wigner's momentum eigenstates in the limit of infinite AdS curvature radius $\ell \to \infty$, at fixed $m=\Delta/\ell$. A simple conformal map relates this description to conformal fields in $\mathbb{R}^3$ familiar from radial quantization, with the interior and exterior of the unit ball respectively corresponding to outgoing and ingoing momenta. Under the contraction, and upon appropriate rescaling, the inner product on $\mathcal{D}(\Delta,s)$ converges to the standard Lorentz-invariant inner product of Wigner's massive particles, so that unitarity is preserved throughout. The contraction to massless particles is briefly discussed here, and will be treated in detail in a forthcoming paper.
		\end{abstract}
		
	\end{titlepage}
	
	{\hypersetup{linkcolor=black}
		\tableofcontents 
	}
	
	\section{Introduction} 
	
	Ever since the emergence of the AdS/CFT correspondence \cite{Aharony:1999ti}, there have been attempts to formulate a \textit{flat limit} whereby conformal correlators are mapped to scattering amplitudes \cite{Polchinski:1999ry,Susskind:1998vk,Giddings:1999qu,Giddings:1999jq}. The physical picture is simple: a scattering process confined to a region much smaller than the AdS curvature radius $\ell$ should not resolve the curvature, and the corresponding S-matrix should be accessible from the boundary data. Turning this heuristic picture into a precise statement has proved subtle, and a variety of prescriptions have been put forward in Mellin space, in position space, in momentum space, and in the axiomatic setting of quantum field theory on a fixed AdS background \cite{Heemskerk:2009pn,Penedones:2010ue,Paulos:2011ie,Fitzpatrick:2010zm,Fitzpatrick:2011ia,Fitzpatrick:2011hu,Fitzpatrick:2011dm,Paulos:2016fap,Komatsu:2020sag,Li:2021snj,Gadde:2022ghy,vanRees:2022zmr,vanRees:2023fcf,Marotta:2024sce,Bagchi:2023fbj,Alday:2024yyj,Lipstein:2025jfj}. These prescriptions are statements about correlation functions. The Hilbert space underlying these correlators usually plays no explicit role, and the fate of the conformal multiplets themselves in the flat limit is left implicit. A detailed understanding of the flat limit at the level of the Hilbert space is valuable both conceptually and for practical reasons, and it is the aim of the present work to provide one.
	
	More precisely, we wish to develop a clear picture of the process by which scattering states, defined as unitary irreducible representations (UIR) of the Poincaré group, emerge from a flat limit of the conformal UIRs. At the level of algebras, it is well-known that this flat limit corresponds to the Inönü--Wigner contraction $\mathfrak{so}(2,3) \to \mathfrak{iso}(1,3) $ \cite{Inonu:1953sp,Saletan:1961iqo,Mickelsson:1972fh}, and general mathematical results about the corresponding contraction of unitary representations have been obtained in \cite{Evans:1967zgs,Angelopoulos:1980wg}. In particular, Evans established that massive Poincaré particles are obtained from a sequence of conformal representations where the scaling dimension $\Delta$ is taken to infinity \cite{Evans:1967zgs}. From the AdS$_4$/CFT$_3$ perspective, this result is easily understood from the standard relation 
	\begin{equation}
		\label{ads dictionary}
		\Delta=\frac{3}{2}+\sqrt{\left(s-\frac{1}{2}\right)^2+m^2 \ell^2} \quad  \stackrel{\ell \to \infty}{\sim} \quad  m \ell\,, \qquad (\text{fixed }m, s\neq 0)
	\end{equation}
	whereby the mass $m$ and spin $s$ of a bulk field are held fixed in the flat limit $\ell \to \infty$, so that the scaling dimension of the dual operator diverges as $\Delta \sim m \ell$. Around the same time, Fronsdal and his collaborators developed the basics of quantum field theory in anti-de Sitter (AdS) space,\footnote{Called \textit{de Sitter space} in that series of work.} including a precise dictionary between free fields of arbitrary mass and spin and lowest-weight representations of the conformal group \cite{Fronsdal:1965zzb,Fronsdal:1974ew,Fronsdal:1975eq,Fronsdal:1975ac,Fang:1978se,Flato:1978qz,Fronsdal:1978vb,Fang:1979hq,Binegar:1981xc}. Their work largely anticipated the AdS/CFT correspondence discovered two decades later \cite{Maldacena:1997re,Gubser:1998bc,Witten:1998qj}, albeit in a non-supersymmetric setting and from the perspective of low-energy `particle physics'. Crucially, Fronsdal realized that unitary lowest-weight representations of $\operatorname{SO}(2,3)$ can be described in a basis of \textit{pseudo-momentum} states that directly reduce to Wigner's momentum eigenstates upon the contraction $\mathfrak{so}(2,3) \to \mathfrak{iso}(1,3)$ \cite{Fronsdal:1965zzb}, and which he described in detail for scalar \cite{Fronsdal:1974ew} and spin-$\frac{1}{2}$ representations \cite{Fronsdal:1975eq}. 
	
	Compared to previous works, the present paper provides the following additions:
	\begin{itemize}
		\item[1.] The construction of the pseudo-momentum basis for lowest-weight representations $\mathcal{D}(\Delta,s)$ with \textit{arbitrary} scaling dimension $\Delta$ and integer spin $s$, including an explicit formula for the corresponding unitary inner product;
		\item[2.] A simple conformal transformation between the pseudo-momentum operators and the Euclidean conformal fields in $\mathbb{R}^3$ familiar from radial quantization, including the observation that outgoing momenta are mapped to the interior of the unit ball while ingoing momenta are mapped to its exterior;
		\item[3.] An explicit description of the flat limit contraction $\ell \to \infty$, whereby the pseudo-momentum states of a lowest-weight representation $\mathcal{D}^{(\ell)}(\Delta,s)$ reduce to Wigner's momentum eigenstates of positive mass $m=\Delta/\ell$ and spin $s$;
		\item[4.] A sequence of unitary inner products $\langle \cdot | \cdot \rangle_\ell$ on $\mathcal{D}^{(\ell)}(\Delta,s)$ converging to the standard unitary inner product for Wigner's particles.  
	\end{itemize}
	The formalism we develop allows straightforward generalization to the higher-dimensional contraction $\mathfrak{so}(2,d) \to \mathfrak{iso}(1,d)$, although we restrict the present discussion to $d=3$ for concreteness. We also restrict to strictly massive particles $m>0$, leaving the contraction to massless particles to the next paper in this series \cite{Nguyen:appear}. 
	
	We stress that we will never refer to local bulk physics in anti-de Sitter space, as a description of the Hilbert space does not require doing so. Interestingly though, the bulk wavefunctions corresponding to the pseudo-momentum states described here have recently been constructed in \cite{Berenstein:2025tts,Berenstein:2025qhb}, albeit without reference to the earlier work of Fronsdal. A related line of work studies the flat limit at the level of bulk local states~\cite{Hao:2025btl,Hao:2026cqm,Hao:2026bop}. There, the lowest-weight conditions are shown to contract to the rest-frame conditions of a massive Poincar\'e particle, in AdS$_3$ for spin~0 and~1 \cite{Hao:2025btl,Hao:2026bop} and in AdS$_{d+1}$ for spin~0 \cite{Hao:2026cqm}. Boosted momentum states are constructed in \cite{Hao:2026cqm} and their Lorentz-invariant normalization is recovered from a sum over descendants at level $n\sim\ell$. The present analysis is
	complementary. It applies to arbitrary integer spin, keeps track of the unitary inner product at finite $\ell$, and works directly in a continuous pseudo-momentum basis which makes the emergence of the Lorentz-invariant inner product manifest. 
	
	The paper is organized as follows. In Section~\ref{section 2} we review the description of unitary lowest-weight representations $\mathcal{D}(\Delta,s)$ of $\operatorname{SO(2,3)}$ in terms of parabolic Verma modules, their realization in terms of Euclidean conformal fields on $\mathbb{R}^3$, and the unitary inner product inherited from radial quantization. In Section~\ref{section 3} we generalize Fronsdal's construction of pseudo-momentum states to a representation $\mathcal{D}(\Delta, s)$ of arbitrary $\Delta$ and integer $s$, including the corresponding inner product worked out in Appendix~\ref{app inner product}. We also describe the conformal map existing between the two descriptions of the same Hilbert space furnished by the conformal fields in $\mathbb{R}^3$ and the pseudo-momentum fields on the mass hyperboloid $\mathbb{H}^3$. We note that the interior of the unit ball maps to future-directed momenta and its exterior to past-directed ones, in accordance with the \textit{all-outgoing} convention found in the scattering literature. A non-trivial identity establishing the equivalence of the two inner products is worked out in Appendix~\ref{app: Wigner rotation}. In Section~\ref{section 4} we implement the flat limit and show that, after an appropriate $\ell$-dependent rescaling, the unitary inner product contracts to the standard Lorentz-invariant inner product for massive particle states, so that unitarity is preserved throughout the flat limit contraction. We conclude in Section~\ref{section 5} with a preliminary discussion of the contraction to massless particles, and with open questions regarding the relevance of our results to \textit{flat-space holography} in general.
	
	\section{Unitary representations of the conformal group}
	\label{section 2}
	This section collects the standard description of the unitary irreducible representations of the conformal group $\operatorname{SO}(2, 3)$, both to fix our conventions and to set up the two realizations of the same Hilbert space that we will use throughout: the abstract lowest-weight modules built from the maximal compact subalgebra, and the Euclidean conformal fields on $\mathbb{R}^3$ providing a state-operator correspondence. A third realization in terms of pseudo-momentum states will be introduced in Section~\ref{section 3}.
	
	We present the conformal algebra $\mathfrak{so}(2,3)$ in the standard form
	\begin{equation}
		\left[L_{AB},L_{CD}\right]=i\left(\eta_{AC} L_{BD}+\eta_{BD}L_{AC}-\eta_{AD} L_{BC}-\eta_{BC} L_{AD} \right)\,,
	\end{equation}
	where the indices run over $A=0,i,4$ with $i=1,2,3$, and the nonzero components of the flat metric tensor $\eta_{AB}$ are given by
	\begin{equation}
		\eta_{00}=\eta_{44}=-1\,, \qquad \eta_{ij}=\delta_{ij}\,.
	\end{equation}
	We will restrict our attention to unitary representations, for which generators are hermitian in the chosen conventions,
	\begin{equation}
		\label{hermiticity condition}
		L_{AB}^\dagger=L_{AB}\,.
	\end{equation}
	
	\subsection{Verma modules}
	We start by reviewing the description of unitary irreducible representations (UIR) of the conformal group $\operatorname{SO}(2,3)$ in terms of parabolic Verma modules as originally developed in \cite{Fronsdal:1965zzb,Evans:1967zgs,Fronsdal:1974ew,Fronsdal:1975eq,Fronsdal:1975ac} and nicely reviewed in \cite{Nicolai:1984hb,deWit:1999ui}. 
	
	The unitary representations of $\operatorname{SO}(2,3)$ are constructed from irreducible representations of its maximal compact subgroup $\operatorname{SO}(2) \times \operatorname{SO}(3)$, respectively generated by $D \equiv L_{40}$ and $L_{ij}$. One thus starts with a lowest-weight multiplet $|\Delta, s h \rangle $ carrying a spin-$s$ representation of $\operatorname{SO}(3)$, i.e.,
	\begin{equation}
		\label{primary states}
		D|\Delta, s h \rangle =\Delta |\Delta, s h \rangle\,, \qquad L_{12} |\Delta, s h \rangle =h |\Delta, s h \rangle\,, 
	\end{equation}
	where $h$ runs over $h=-s,-s+1,...,s$. For integer $s$, we can encode the same spin-$s$ multiplet into a totally symmetric traceless tensor of rank $s$,
	\begin{equation}
		\{|\Delta, sh\rangle\}_{h=-s,...,s} \qquad \longleftrightarrow \qquad |\Delta\rangle_I\equiv |\Delta\rangle_{i_1...\,i_s}\,,
	\end{equation}
	where we introduced the multi-index $I=i_1...\,i_s$. This second encoding is better suited for generalizations to unitary representations of $\operatorname{SO}(2,d)$ with arbitrary dimension $d$.
	Within this lowest-weight multiplet, the Casimir operator of the rotation group $\vec L^2=\frac{1}{2} L_{ij}L^{ij}$ takes the constant value $s (s +1)$. The remaining generators are arranged into raising and lowering operators,
	\begin{equation}
		\label{ladder operators}
		L_i^\pm =L_{i4}\mp i L_{i0} \,,
	\end{equation}
	that satisfy the commutation relations
	\begin{align}
		\left[D, L^\pm_i \right]=\pm L^\pm_i\,, \qquad \left[L^\pm_i, L^\pm_j \right]=0\,, \qquad \left[L_i^+, L_j^- \right]=-2 (\delta_{ij} D+i L_{ij})\,,
	\end{align}
	and are conjugate to each other $(L^\pm_i)^\dagger=L_i^\mp$. Since $|\Delta, s h \rangle$ are assumed to be of lowest-weight, we must have
	\begin{equation}
		\label{primary condition}
		L^-_i |\Delta, s h \rangle =0\,.
	\end{equation}
	The rest of the representation is constructed by acting with the raising operators $L^+_i$. Acting with an $n$-fold product of the raising operators, one obtains descendant states with weight $\Delta+n$. The full Verma module $\mathcal{V}_{\Delta,s}$ may be expressed as a sum over the levels $n$,
	\begin{equation}
		\mathcal{V}_{\Delta,s}\simeq \bigoplus_{n=0}^\infty\mathcal{V}_{\Delta,s,n}\,, \qquad \mathcal{V}_{\Delta,s,n}=\text{span}\{L^+_{i_1}...\,L^+_{i_n}\, |\Delta, s h \rangle \}\,.
	\end{equation}
	Each level has definite weight $\Delta+n$ and can be organized into states of definite angular momentum. This amounts to a decomposition of the following tensor product of $\operatorname{SO}(3)$ irreps,
	\begin{equation}
		\label{level n tensor product}
		\underbrace{1 \otimes ... \otimes 1}_{n\, \text{times}} \otimes\, s\,,
	\end{equation}
	where the first $n$ occurrences of the vector representation result from the $n$-fold product $L^+_{i_1}...\,L^+_{i_n}$, and the last factor comes from the spin-$s$ representation carried by the lowest-weight multiplet.
	
	Useful information is obtained by considering the value of the quadratic Casimir operator within the lowest-weight representation. This operator is given by
	\begin{equation}
		\label{Casimir}
		\begin{split}
			\mathcal{C}_2^{\mathfrak{so}(2,3)}&=\frac{1}{2} L_{AB}L^{AB}=D^2-\frac{1}{2}\delta^{ij}\{ L_i^+\,, L_j^- \}+\vec L^2\\
			&=D(D-3)-\delta^{ij} L_i^+ L_j^- +\vec L^2\,.
		\end{split}
	\end{equation}
	Acting on the ground states, one finds 
	\begin{equation}
		\label{constant Casimir value}
		\mathcal{C}_2^{\mathfrak{so}(2,3)}=\Delta (\Delta-3)+s (s+1)\,,
	\end{equation}
	which is therefore its value on any state contained in $\mathcal{V}_{\Delta,s}$. One can act with the Casimir operator on a descendant state at level $n=1$. Assuming that $s \geq 1$, the decomposition of the tensor product \eqref{level n tensor product} at level $n=1$ is simply
	\begin{equation}
		1 \otimes s = (s-1) \oplus s \oplus (s +1)\,.
	\end{equation}
	Acting on a level-1 state $|\Psi \rangle$ of spin $(s-1)$, 
	\begin{equation}
		\label{normalized level 1}
		|\Psi \rangle \equiv  \sum_i L_i^+ |\Delta \rangle_{ii_2...\,i_s}\,, 
	\end{equation}
	we thus get
	\begin{equation}
		\langle \Psi|\mathcal{C}_2^{\mathfrak{so}(2,3)}|\Psi\rangle=\langle \Psi|\Psi\rangle\, [(\Delta+1)(\Delta-2)+(s-1)s]-\sum_i \|L_i^-|\Psi\rangle\|^2\,,
	\end{equation}
	where the appearance of the norm follows from the conjugation relation $(L^+_i)^\dagger=L^-_i$. To belong to the same irrep as the ground states, this must equal \eqref{constant Casimir value}, which yields the condition
	\begin{equation}
		\label{norm of descendant}
		\sum_i \|L_i^-|\Psi \rangle\|^2=2(\Delta-s-1)\langle \Psi|\Psi \rangle\,.
	\end{equation}
	Thus, unitarity of the representation and positivity of the norm together yield a lower bound on the scaling dimension,
	\begin{equation}
		\Delta \geq s+1\,, \qquad (s\geq 1)\,.
	\end{equation}
	It can be shown that this condition is sufficient to guarantee unitarity of $\mathcal{V}_{\Delta,s}$. For $\Delta > s+1$, it is also irreducible.  When $\Delta=s+1$, the norm \eqref{norm of descendant} vanishes and the Verma module $\mathcal{V}_{s+1,s}$ is actually reducible. The level-1 states \eqref{normalized level 1} are the ground states of another lowest-weight representation $\mathcal{V}_{s+2,s-1} \subset \mathcal{V}_{s+1,s}$. A unitary irreducible representation is obtained by taking the quotient of these two modules,
	\begin{equation}
		\mathcal{D}(s+1,s)=\mathcal{V}_{s+1,s}/\mathcal{V}_{s+2,s-1}\,, \qquad (s \geq 1)\,,
	\end{equation}
	i.e., by removing the states \eqref{normalized level 1} and their descendants. This is implemented by the null state/shortening condition
	\begin{equation}
		\label{null state}
		\sum_i L_i^+ |\Delta\rangle_{ii_2...i_s}=0\,, \qquad \forall\, i_2,...\,i_s\,, \qquad (\Delta=s+1)\,.
	\end{equation}
	
	The case $s=0$ must be treated separately \cite{Fronsdal:1974ew}. The unitarity bound is established from level-2 states and yields $\Delta \geq 1/2$. As with spinning representations, the module $\mathcal{V}_{\frac{1}{2},0}$ that saturates this bound is reducible. Its irreducible component is Dirac's singleton representation \cite{Dirac:1963ta}. 
	
	\subsection{State-operator correspondence}
	There exists a simple correspondence between states of the Verma modules as described above and local operators on the Euclidean\footnote{We stress that the correspondence is with Euclidean rather than Lorentzian tensor fields. Indeed, the subgroup $\operatorname{SO}(3)$ of the maximal compact subgroup $\operatorname{SO}(2) \times \operatorname{SO}(3)$, which is identified with the rotation group in $\mathbb{R}^3$, acts unitarily on the corresponding tensor representations. This is a necessary condition for a correspondence with the unitary modules $\mathcal{V}_{\Delta,s}$.} space $\mathbb{R}^3$. By definition, the ground states \eqref{primary states} of the module $\mathcal{V}_{\Delta,s}$ correspond to the independent components of a totally symmetric and traceless tensor operator of rank $s$ and scaling dimension $\Delta$, placed at the origin of $\mathbb{R}^3$,
	\begin{equation}
		|\Delta \rangle_{I} =O^\Delta_{I}(0)|0\rangle\,, \qquad I=i_1...\,i_s\,,
	\end{equation}
	where the vacuum state transforms in the trivial representation. The rest of the module is generated by acting with the exponentiated ladder operators $L_i^+$, 
	\begin{equation}
		\label{local conformal fields}
		O^\Delta_I(\vec x)=e^{x^i L_i^+}\, O^\Delta_I(0)\, e^{-x^i L_i^+}\,, \qquad x<1 \,,
	\end{equation}
	where $x=\sqrt{x^i x_i}$. The restriction to the interior of the unit ball ensures that the corresponding state has finite norm (see \eqref{norm O(x)} below).
	Using the algebra relations, it can be shown that the $\mathfrak{so}(2,3)$ generators act according to \cite{Mack:1969rr}
	\begin{equation}
		\begin{split}
			\left[L_i^+, O^\Delta_I(\vec x)\right]&=\partial_i O^\Delta_I(\vec x)\,,\\
			\left[L_i^-, O^\Delta_I(\vec x)\right]&=(2\Delta x_i+2x_i x^j \partial_j-x^2 \partial_i-2i x^j[L_{ij}]^{(s)} )\,O^\Delta_I(\vec x)\,,\\
			\left[D, O^\Delta_I(\vec x)\right]&=(\Delta+x^i\partial_i )\,O^\Delta_I(\vec x)\,,\\
			\left[L_{ij}, O^\Delta_I(\vec x)\right]&=([L_{ij}]^{(s)}+i(x_i \partial_j-x_j\partial_i))\,O^\Delta_I(\vec x)\,,
		\end{split}
	\end{equation}
	where $[L_{ij}]^{(s)}$ is the rank-$s$ totally symmetric and traceless tensor representation of $L_{ij}$.
	The raising operators $L_i^+$ (often denoted $P_i$) are seen to generate translations in $\mathbb{R}^3$ as follows directly from the definition \eqref{local conformal fields}. Hence, level-$n$ descendant states in $\mathcal{V}_{\Delta,s}$ correspond to $n$-fold spatial derivatives of the local tensor operator evaluated at the origin.
	
	In this language of Euclidean tensor fields, the shortening condition \eqref{null state} simply reads
	\begin{equation}
		\partial^i O_{ii_2...\,i_s}(0)=0\,, \qquad (\Delta=s+1)\,.
	\end{equation}
	Since $L_i^+$ trivially commutes with the translation element $e^{x^i L_i^+}\in \operatorname{SO}(2,3)$, the same condition also holds everywhere in the unit ball,
	\begin{equation}
		\label{conservation equation}
		\partial^i O_{ii_2...\,i_s}(\vec x)=0\,, \qquad \forall \, x<1\,, \qquad (\Delta=s+1)\,.
	\end{equation}
	For that reason, the operator $O_I=O_{i_1...\, i_s}$ corresponding to a short multiplet is often called a \textit{conserved current}.
	
	\subsection{Inner product and reflection positivity}
	The inner product on the module $\mathcal{V}_{\Delta,s}$ is inherited from the $\operatorname{SO}(3)$-invariant inner product for the lowest-weight multiplet,
	\begin{equation}
		\label{lowest inner product}
		{}_I\langle \Delta|\Delta \rangle_J=\delta_{IJ}\,,
	\end{equation}
	where $\delta_{IJ}$ is the projector onto traceless symmetric tensors.
	The inner product of any two descendant states may be derived from it using the algebra relations.
	
	In practice, it is often more economical to work in terms of the Euclidean tensor fields and their correlation functions. Conformal Ward identities uniquely fix the form of the two-point function to be \cite{Osborn:1993cr,Costa:2011mg}
	\begin{equation}
		\label{CFT 2-point function}
		\langle 0|O_I(\vec x_1) O_J(\vec x_2) |0\rangle=\frac{\mathcal{I}
			_{IJ}(\vec x_{12})}{|\vec x_{12}|^{2\Delta}}\,, \qquad \vec x_{12}\equiv \vec x_1-\vec x_2\,,
	\end{equation}
	where $\mathcal{I}_{IJ}(\vec x)$ is the spin-$s$ inversion tensor, constructed as the totally symmetric and traceless part of the $s$-fold product of
	\begin{equation}
		\mathcal{I}_{ij}(\vec x)=\delta_{ij}-\frac{2 x_i x_j}{x^2}\,.
	\end{equation}
	As already mentioned, hermiticity of the $\mathfrak{so}(2,3)$ generators \eqref{hermiticity condition} implies that the ladder operators \eqref{ladder operators} are conjugate to each other, $(L_i^\pm)^\dagger=L_i^\mp$. As is well-known, this operation may be implemented in $\mathbb{R}^3$ through inversion $\mathcal{R}: \vec x \mapsto \vec x/x^2$,
	\begin{equation}
		\mathcal{R} L_i^\pm \mathcal{R}=L_i^\mp\,.
	\end{equation}
	This implies that the state conjugated to $|O_I(\vec x)\rangle=O_I(\vec x)|0\rangle$ is $\langle O_I(\vec x)|=\langle 0| [O_I(\vec x)]^\dagger$ with \cite{Rychkov:2016iqz,Osborn:CFTNotes}
	\begin{equation}
		\label{conjugate operator}
		[O_I(\vec x)]^\dagger=\mathcal{R} O_I(\vec x) \mathcal{R}=x^{-2\Delta}\, \mathcal{I}\indices{_I^J}(\vec x)\, O_J(\vec x/x^2)\,.
	\end{equation} 
	Since normalizable states (kets) correspond to operator insertions inside the unit ball, conjugate states (bras) correspond to operator insertions outside the unit ball.
	The overlap of two states then follows, using \eqref{CFT 2-point function},
	\begin{equation}
		\label{state overlap}
		\begin{split}
			\langle O_I(\vec x_1)|O_J(\vec x_2) \rangle&=x_1^{-2\Delta}\, \mathcal{I}\indices{_I^K}(\vec x_1) \langle 0| O_K(\vec x_1/x_1^2) O_J(\vec x_2) |0\rangle\\
			&=\frac{\mathcal{I}\indices{_I^K}(\vec x_1)\, \mathcal{I}
				_{KJ}(\vec x_1/x_1^2-\vec x_2)}{(1-2\, \vec x_1 \cdot \vec x_2+x_1^2\, x_2^2)^\Delta}\,.
		\end{split}
	\end{equation}
	This overlap defines a positive-definite inner product provided the spectrum satisfies the unitarity bounds stated earlier. Thus, Hilbert space unitarity implies \textit{reflection positivity} of two-point Euclidean correlators, a particular instance of the Osterwalder--Schrader theory \cite{Osterwalder:1973dx,Osterwalder:1974tc} discussed in \cite{Kravchuk:2021kwe}.
	At coincident points, \eqref{state overlap} becomes
	\begin{equation}
		\label{norm O(x)}
		\langle O_I(\vec x)|O_J(\vec x) \rangle=\frac{\mathcal{I}\indices{_I^K}(\vec x)\, \mathcal{I}
			_{KJ}(\vec x)}{(1-x^2)^{2\Delta}}=\frac{\delta_{IJ}}{(1-x^2)^{2\Delta}}>0\,, \qquad (x<1)\,,
	\end{equation}
	where we made use of the property $\mathcal{I}_{IJ}(f \vec x)=\mathcal{I}_{IJ}(\vec x)$, valid for any non-vanishing function $f$, and of $\mathcal{I}\indices{_I^K}(\vec x)\mathcal{I}_{KJ}(\vec x)=\delta_{IJ}$. For $\vec x=0$, \eqref{norm O(x)} reduces back to the inner product \eqref{lowest inner product} of the lowest-weight multiplet.
	
	\section{The pseudo-momentum basis}
	\label{section 3}
	In order to describe the contraction of the lowest-weight representations in the flat limit, it will be useful to introduce yet another description in terms of \textit{pseudo-momentum states} as originally done by Fronsdal for the scalar UIR $\mathcal{D}(\Delta,0)$ \cite{Fronsdal:1974ew}, and by Fronsdal and Haugen for the spin-$\frac{1}{2}$ UIR $\mathcal{D}(\Delta,\frac{1}{2})$ \cite{Fronsdal:1975eq}. In this section, we generalize this construction to arbitrary integer half-spin $s \in \frac{1}{2}\mathbb{N}$.  
	
	\subsection{Definition}
	Following \cite{Fronsdal:1965zzb,Fronsdal:1974ew}, we introduce a new basis of generators of the conformal algebra $\mathfrak{so}(2,3)$, namely
	\begin{equation}
		\label{second generator basis}
		L_{\mu\nu}\,, \qquad \mathcal P_\mu\equiv c\, L_{\mu 4}\,, \qquad \mu,\nu=0,1,2,3\,,
	\end{equation}
	satisfying the commutation relations
	\begin{equation}
		\label{conformal algebra momentum basis}
		\begin{split}
			\left[L_{\mu\nu},L_{\rho\sigma}\right]&=i\left(\eta_{\mu \rho} L_{\nu \sigma}+\eta_{\nu\sigma}L_{\mu\rho}-\eta_{\mu\sigma} L_{\nu \rho}-\eta_{\nu\rho} L_{\mu\sigma} \right)\,,\\
			\left[L_{\mu\nu},P_\rho \right]&=i \left(\eta_{\mu\rho}\mathcal P_\nu-\eta_{\nu\rho}\mathcal P_\mu \right)\,,\\
			\left[\mathcal P_\mu,\mathcal P_\nu \right]&=-i c^2 L_{\mu\nu}\,.
		\end{split}
	\end{equation}
	The parameter $c>0$ is introduced in order to implement the Inönü--Wigner contraction $\mathfrak{so}(2,3) \to \mathfrak{iso}(1,3)$ in the limit $c \to 0^+$. Indeed, in that limit the \textit{pseudo-momentum} generators $\mathcal P_\mu$ commute and may be identified with the standard momentum generators of the Poincaré algebra. Armed with this observation, it is useful to revisit the lowest-weight representations in a way which naturally  connects to Wigner's construction of massive Poincaré particles \cite{Wigner:1939cj}. 
	
	First, we recognize that the dilation operator is also the pseudo-energy operator, $\mathcal P^0=c\, D$, such that the equations \eqref{primary states} satisfied by primary states are similar to those defining spin-$s$ massive Poincaré particles in their rest frame, i.e.,
	\begin{equation}
		\label{primary conditions bis}
		\mathcal P_0|\Delta, s h\rangle=-m|\Delta, s h\rangle\,, \qquad L_{12} |\Delta, s h\rangle=h|\Delta, s h\rangle\,, 
	\end{equation}
	where the rest mass is identified with
	\begin{equation}
		\label{m=cD}
		m=c\, \Delta>0\,.
	\end{equation}
	Similarly to lowest-weight representations, massive Poincaré representations are labeled by their rest energy and $\operatorname{SO}(3)$ spin. Of course, the correspondence can only become exact in the limit $c \to 0$ where the algebra \eqref{conformal algebra momentum basis} contracts to the Poincaré algebra. To keep the mass \eqref{m=cD} finite in that limit will require scaling $\Delta=c^{-1} m \to \infty$ simultaneously. Thus, the parameter $c$ may be identified with the inverse curvature radius $\ell^{-1}$ appearing in \eqref{ads dictionary}. Because of this correspondence, it is useful to define \textit{pseudo-momentum states} in $\mathcal{V}_{\Delta,s}$ which would become eigenstates of $\mathcal P_\mu$ in the limit $c \to 0$. Following Wigner's construction, one therefore introduces the standard boost $L(\hat p)$ that maps the rest momentum $k^\mu=m\, \delta^\mu_0$ to a generic momentum $p^\mu$ on the orbit $p^2=-m^2$, and thus satisfies
	\begin{equation}
		\label{boost matrix}
		\hat p^\mu=L(\hat p)\indices{^\mu_0}\,, \qquad \hat p^\mu\equiv p^\mu/m\,.
	\end{equation}
	The explicit expression of $L(\hat p)\indices{^\mu_\nu}$ is given in \eqref{boost matrix}. The pseudo-momentum states are then defined from the primary states through \cite{Fronsdal:1965zzb,Fronsdal:1974ew}
	\begin{equation}
		\label{pseudo momentum state}
		|p, sh \rangle \equiv U(L(\hat p)) |\Delta, sh \rangle\,.
	\end{equation}
	Under a Lorentz transformation $\Lambda \in \operatorname{SO}(1,3)$, a standard argument yields the transformation rule \cite{Weinberg:1995mt}
	\begin{equation}
		\label{Lorentz transfo}
		U(\Lambda) |p,sh\rangle = \sum_{h'} [W(\Lambda,p)]_{h'h} |\Lambda p, sh'\rangle\,,
	\end{equation}
	where $W(\Lambda,p)$ is an element of the little group $\operatorname{SO}(3)$ leaving $k^\mu=m\, \delta^\mu_0$ invariant, given explicitly by
	\begin{equation}
		\label{Wigner rotation abstract}
		W(\Lambda,p)\equiv L^{-1}(\Lambda \hat p) \Lambda L(\hat p)\,,
	\end{equation}
	and $[W]_{h'h}=\langle sh'|U(W)|sh\rangle$ the corresponding spin-$s$ representation in the canonical basis.\footnote{The spin projection $h$ does not coincide with the helicity, as we do not adopt the convention of Jacob and Wick \cite{Jacob:1959at} for the standard boost $L(\hat p)$.} This is exactly identical to the Lorentz transformation of a massive Poincaré particle. Infinitesimally, this reads \cite{Fronsdal:1965zzb}\footnote{Details of this derivation may be found for instance in \cite{Have:2024dff}; see (3.15) therein with $J_{ab} \to -i L_{ij}$ and $B_a \to -i L_{0i}$.}
	\begin{subequations}
		\begin{align}
			L_{ij} |p,sh\rangle&=\left[ [L_{ij}]^{(s)}+i \left(\hat p_i \frac{\partial}{\partial \hat p^j}-\hat p_j \frac{\partial}{\partial \hat p^i}\right)\right]|p,sh\rangle\,,\\
			L_{0i}|p,sh\rangle&=\left[ \frac{\hat p^j}{\hat p^0+1} [L_{ij}]^{(s)}+i \left(\hat p_0 \frac{\partial}{\partial \hat p^i}-\hat p_i \frac{\partial}{\partial \hat p^0}\right)\right]|p,sh\rangle\,,
		\end{align}
	\end{subequations}
	where $[L_{ij}]^{(s)}$ is the spin-$s$ representation of the rotation generator $L_{ij}$. The action of the pseudo-momentum generator $\mathcal P_\mu$ in the rest frame is obtained from \eqref{primary conditions bis} and the lowest-weight condition \eqref{primary condition}, and can be collected into the single equation
	\begin{equation}
		\mathcal P_\mu |\Delta,sh\rangle=\left[k_\mu+i c\, \hat k^\nu L_{\nu \mu} \right] |\Delta,sh\rangle\,, \qquad \hat k^\mu=k^\mu/m=\delta^\mu_0\,.
	\end{equation}
	The action of $\mathcal P_\mu$ on a generic pseudo-momentum state \eqref{pseudo momentum state} is simply obtained by replacing $k^\mu \to p^\mu$, 
	\begin{equation}
		\label{P action}
		\mathcal P_\mu |p,sh\rangle=\left[p_\mu+i c\, \hat p^\nu L_{\nu \mu} \right] |p,sh\rangle\,.
	\end{equation}
	Hence pseudo-momentum states do not diagonalize the pseudo-momentum generator $\mathcal P_\mu=c L_{\mu 4}$, which is a manifestation of the nonzero commutator $[\mathcal P_\mu, \mathcal P_\nu]=-i c^2 L_{\mu\nu}$. In the limit $c \to 0$ at fixed $p_\mu$, this commutator vanishes and $\mathcal P_\mu$ becomes diagonal with eigenvalue $p_\mu$.
	
	The inner product of the pseudo-momentum states, within the unitary representation $\mathcal{D}(\Delta,s)$, is given by 
	\begin{equation}
		\label{pseudo-momentum kernel}
		\langle p_1, s h_1 | p_2, s h_2 \rangle =  \left(\frac{2}{1-\hat p_1 \cdot \hat p_2} \right)^{\Delta}\, [W_{12}]_{h_1h_2}\,,
	\end{equation}
	where $W_{12}=W(L^{-1}(\hat p_1),p_2)$ is a particular instance of the Wigner rotation \eqref{Wigner rotation abstract} around the axis $\vec p_1 \times \vec p_2$ (see Appendix~\ref{app: Wigner rotation}). See Appendix~\ref{app inner product} for the derivation of \eqref{pseudo-momentum kernel}, which generalizes the expression found in \cite{Fronsdal:1974ew,Fronsdal:1975eq} to arbitrary half-integer spin $s$. Note that the scalar inner product has also been derived from the perspective of wavefunctions in AdS in \cite{Berenstein:2025tts} (equation (40) therein).
	
	\subsection{From position to pseudo-momentum space}
	We have at our disposal two different descriptions of the Hilbert space furnishing the carrier space of the representations $\mathcal{D}(\Delta,s)$: the Euclidean conformal fields in $\mathbb{R}^3$ familiar from radial quantization, and the fields on the pseudo-momentum mass-shell $\mathbb{H}^3$. A simple guess, which turns out to be correct, is that they are related by the conformal mapping between $\mathbb{R}^3$ and $\mathbb{H}^3$, namely
	\begin{equation}
		\label{Psi(p) from O(x)}
		\Psi_I(\hat p(\vec x))=|1-x^2|^{\Delta}\, O_I(\vec x)\,,
	\end{equation}
	together with
	\begin{equation}
		\label{flat momentum param}
		\hat p^\mu(\vec x)=\left(\frac{1+x^2}{1-x^2}\,, \frac{2 \vec x}{1-x^2} \right)\,.
	\end{equation}
	Indeed, the metric on $\mathbb{H}^3$ in the coordinate system $x^a=(x^1, x^2, x^3)$ defined through \eqref{flat momentum param} is explicitly conformally flat,
	\begin{equation}
		g_{ab}=\omega(x)^2\, \delta_{ab}\,, \qquad \omega(x)=\frac{2}{1-x^2}\,.
	\end{equation}
	We will show that \eqref{Psi(p) from O(x)}-\eqref{flat momentum param} provide the correct relation by re-deriving the pseudo-momentum inner product \eqref{pseudo-momentum kernel}. But let us first pause to make an important observation: the interior of the unit ball ($x<1$) corresponds to future-directed momenta ($\hat p^0 >0$), while the exterior of the unit ball ($x>1$) corresponds to past-directed momenta ($\hat p^0<0$). This precisely agrees with the \textit{all-outgoing} convention of scattering theory where incoming particles are labeled with past-directed momenta. Furthermore, inversion $\mathcal{R}: \vec x \mapsto \vec x/x^2$ has the simple action $\hat p^\mu \mapsto -\hat p^\mu$. 
	
	Now let us demonstrate that we recover the overlap of pseudo-momentum states given in \eqref{pseudo-momentum kernel} from the inner product \eqref{state overlap},
	\begin{equation}
		\begin{split}
			{}_I \langle p_1|p_2\rangle_J &\equiv \langle 0| \Psi_I(\hat p_1)^\dagger\, \Psi_J(\hat p_2) |0\rangle=(1-x_1^2)^\Delta\, (1-x_2^2)^\Delta\, \langle 0| O_I(\vec x_1)^\dagger\, O_J(\vec x_2) |0\rangle\\
			&=\left( \frac{(1-x_1^2)(1-x_2^2)}{1-2 \vec x_1 \cdot \vec x_2+x_1^2\, x_2^2}\right)^\Delta \mathcal{I}\indices{_I^K}(\vec x_1)\, \mathcal{I}
			_{KJ}(\vec x_1/x_1^2-\vec x_2)\,.
		\end{split}
	\end{equation}
	Here, both momenta $p_1, p_2$ are future-directed, such that $\vec x_1, \vec x_2$ are inside the unit ball. The first factor is easily seen to be the scalar part of the pseudo-momentum kernel \eqref{pseudo-momentum kernel} when $\hat p_1=\hat p(\vec x_1)$ and $\hat p_2=\hat p(\vec x_2)$ are parametrized as in \eqref{flat momentum param},
	\begin{equation}
		\left( \frac{(1-x_1^2)(1-x_2^2)}{1-2 \vec x_1 \cdot \vec x_2+x_1^2\, x_2^2}\right)^\Delta=\left( \frac{2}{1-\hat p_1(\vec x_1) \cdot \hat p_2(\vec x_2)}\right)^\Delta\,.
	\end{equation}
	The second factor precisely coincides with the Wigner rotation $[W_{12}]_{IJ}$ appearing in \eqref{pseudo-momentum kernel},
	\begin{equation}
		\label{II=W main}
		\mathcal{I}\indices{_I^K}(\vec x_1)\, \mathcal{I}
		_{KJ}(\vec x_1/x_1^2-\vec x_2)=[W_{12}]_{IJ}\,.
	\end{equation}
	The technical proof of this non-trivial identity is given in Appendix~\ref{app: Wigner rotation}.
	
	\paragraph{Shortening condition.} Although we do not make direct use of it in the present work, for later reference we translate the shortening condition \eqref{null state} to the pseudo-momentum description.
	Inverting the relation \eqref{Psi(p) from O(x)},
	\begin{equation}
		O_I(\vec x)=|1-x^2|^{-\Delta}\, \Psi_I(\hat p(\vec x))\,,
	\end{equation}
	and taking its divergence yields
	\begin{equation}
		\partial^i O_{i i_2...\,i_s}(\vec x)=|1-x^2|^{-\Delta} \left( \partial^i \Psi_{ii_2...\,i_s}(\vec x)+\Delta\, \hat p^i(\vec x)\, \Psi_{ii_2...\,i_s}(\vec x) \right)\,.
	\end{equation}
	Hence, the current conservation \eqref{conservation equation}, where $\Delta=s+1$, is equivalent to 
	\begin{equation}
		\label{divergence Psi}
		\partial^i \Psi_{ii_2...\,i_s}(\vec x)+(s+1)\, \hat p^i(\vec x)\, \Psi_{ii_2...\,i_s}(\vec x) =0\,.
	\end{equation}
	This is actually recognized as a covariant divergence-free condition on the unit mass hyperboloid $\mathbb{H}^3$, namely
	\begin{equation}
		\label{nabla Psi}
		\nabla^a \Psi_{a a_2...\, a_s}=0\,,
	\end{equation}
	where the coordinate components $\Psi_{a_1...\,a_s}$ are related to the frame components $\Psi_{i_1...\,i_s}$ via
	\begin{equation}
		\Psi_{a_1...a_s}=e^{i_1}_{a_1}...\, e^{i_s}_{a_s} \Psi_{i_1...\,i_s}\,,
	\end{equation}
	in terms of the tetrad 
	\begin{equation}
		e^i_a(x)=\omega(x) \delta^i_a\,, \qquad g_{ab}=e_a^i\, \delta_{ij}\, e^j_b\,.
	\end{equation}
	Using this and the Christoffel symbols
	\begin{equation}
		\Gamma^c_{ab}=\delta^c_a\, \hat p_b+\delta^c_b\, \hat p_a-\delta_{ab}\, \hat p^c\,, \qquad \hat p^a(\vec x)=\frac{2 x^a}{1-x^2}\,,
	\end{equation}
	it is easily checked that \eqref{nabla Psi} reduces to \eqref{divergence Psi}.
	
	\section{Massive particles in the flat limit}
	\label{section 4}
	
	We are now ready to implement the flat limit contraction $c \to 0$ by which massive spin-$s$ Poincaré particles are obtained from lowest-weight representations $\mathcal{D}(\Delta,s)$. Most of this contraction has already been anticipated in the previous section, where we introduced the pseudo-momentum basis. Under the Lorentz subgroup, these states transform just like Wigner's massive particles; see \eqref{Lorentz transfo}. As we implement the algebra contraction $c \to 0$, they also become eigenstates of the momentum operators $\mathcal P_\mu$; see \eqref{P action}. Thus, they become indistinguishable from Wigner's massive Poincaré particles. The only thing left to discuss is the fate of the inner product on the Hilbert space as we implement this contraction. 
	
	The standard inner product of massive Poincaré particles is recovered in the $c \to 0$ limit at fixed $p^\mu$ in the following way, starting from \eqref{pseudo-momentum kernel}, which we rescale by some $c$-dependent constant $\mathcal{N}(\Delta,c)>0$,
	\begin{equation}
		\label{pseudo-momentum kernel bis}
		\langle p_1, s h_1 | p_2, s h_2 \rangle_c = \mathcal{N}(\Delta,c) \left(\frac{2}{1-\hat p_1 \cdot \hat p_2} \right)^{\Delta}\, [W_{12}]_{h_1h_2}\,.
	\end{equation}
	When $p_1 \neq p_2$, we have $1-\hat p_1 \cdot \hat p_2 > 2$  and the inner product vanishes in the limit $\Delta=c^{-1} m \to \infty$. Thus, in that limit the distribution \eqref{pseudo-momentum kernel bis} is fully supported at $p_1 = p_2$, where the quantity inside the parentheses equals 1 and the Wigner rotation matrix becomes the identity, $[W_{12}]_{h_1h_2} \to \delta_{h_1 h_2}$. More precisely, we obtain
	\begin{equation}
		\label{kernel in the flat limit}
		\langle p_1, s h_1 | p_2, s h_2 \rangle_c \approx \mathcal{N}(\Delta,c)\, \frac{m^{1/2} c^{3/2}}{2\pi^{3/2}}\, (2\pi)^3\, 2p^0 \delta^{(3)}(\vec p_1-\vec p_2) \delta_{h_1 h_2}\,, \qquad (c \to 0)\,.  
	\end{equation}
	The  prefactor may be computed in the following way. First, let us parametrize the momentum $p_1^\mu$ as
	\begin{equation}
		\label{momentum rapidity param}
		p_1^\mu=m\, (\cosh \eta, \sinh \eta\, \vec n)\,, \qquad \vec n^2=1\,,
	\end{equation}
	where $\eta$ is the rapidity and $\vec n$ is a unit spatial vector. In these coordinates, the invariant volume element is given by 
	\begin{equation}
		[d^3 p_1]\equiv \frac{1}{(2\pi)^3} \frac{d^3 \vec p_1}{2 p_1^0}=\frac{m^2}{2(2\pi)^3} \sinh^2 \eta \dd \eta \dd \Omega\,.
	\end{equation}
	Without loss of generality, we may choose a frame where $ p_2^\mu=(m,\vec 0)$, such that 
	\begin{equation}
		p_1 \cdot  p_2=-m^2\cosh \eta\,.
	\end{equation}
	Hence, we may compute
	\begin{equation}
		\begin{split}
			\int [d^3 p_1]\, \left(\frac{2}{1-\hat p_1 \cdot \hat p_2} \right)^\Delta&=\frac{m^2}{(2\pi)^2} \int_0^\infty d\eta\, \sinh^2 \eta\, \cosh^{-2\Delta} \left(\frac{\eta}{2}\right)\\
			&=\frac{m^2\, \Gamma[\Delta-2]}{2\pi^{3/2} \Gamma[\Delta-1/2]}\approx \frac{m^{1/2} c^{3/2}}{2\pi^{3/2}}\,,
		\end{split}
	\end{equation}
	where in the last approximation we used $\Gamma[\Delta-2]/\Gamma[\Delta-1/2]\approx \Delta^{-3/2}$ in the limit $\Delta \to \infty$ and substituted $\Delta=c^{-1} m$.
	
	Therefore, the standard inner product on the Hilbert space of massive Poincaré particles of mass $m$ and spin $s$ is recovered in the limit $c \to 0$, provided we choose the following $c$-dependent normalization coefficient 
	\begin{equation}
		\mathcal{N}(\Delta,c)=\frac{2\pi^{3/2}}{m^{1/2}c^{3/2}}\,,
	\end{equation}
	such that
	\begin{equation}
		\label{limit inner product}
		\lim_{c \to 0}\, \langle p_1, s h_1 | p_2, s h_2 \rangle_c= (2\pi)^3\, 2p^0 \delta^{(3)}(\vec p_1-\vec p_2) \delta_{h_1 h_2}\,.
	\end{equation}
	This completes the contraction: the one-particle Hilbert space carried by $\mathcal{D}(\Delta,s)$ becomes, in the limit $c \to 0$, the Hilbert space of Wigner's massive particles of mass $m$ and spin $s$, inner product included.
	
	\section{Discussion}
	\label{section 5}
	The statement that a massive particle UIR of mass $m$ and integer spin $s$ can be obtained from a sequence of lowest-weight representations $\mathcal{D}^{(\ell)}(\Delta,s)$, corresponding to the Inönü--Wigner contraction $\mathfrak{so}(2,3) \to \mathfrak{iso}(1,3)$ in a limit $\ell \to \infty$ with fixed $m=\Delta/\ell$, is an old result due to Evans \cite{Evans:1967zgs}. In this work, we provided an explicit description of this limiting process in terms of pseudo-momentum states that reduce to Wigner's momentum eigenstates in the limit $\ell \to \infty$, extending the spin $s=0$ treatments of Fronsdal \cite{Fronsdal:1965zzb,Fronsdal:1974ew}. In doing so, we also established a simple conformal map between Euclidean conformal fields in $\mathbb{R}^3$ familiar from radial quantization, and the pseudo-momentum operators on the mass shell $\mathbb{H}^3$. This should be of high practical value for the study of the flat limit of AdS/CFT, whether at the level of correlators and amplitudes, or at a more structural level. For instance, the identity \eqref{II=W main} would have been difficult to guess without it. 
	
	A natural continuation of the present work is to study the reduction of conformal correlators to massive scattering amplitudes. The framework presented here suggests the following methodology: 
	\begin{itemize}
		\item[1.] Start from conformal correlators in $\mathbb{R}^3$ with points inside and outside the unit ball;
		\item[2.] Implement the conformal transformation \eqref{Psi(p) from O(x)} to the unit mass shell $\mathbb{H}^3$; 
		\item[3.] Take the flat limit $\ell \to \infty$ with $m_i=\Delta_i/\ell$ fixed, resulting in S-matrix elements with all-outgoing momentum convention.
	\end{itemize}
	This prescription, which is directly rooted in the Hilbert space contraction described in this work, differs from most of those already existing in the literature \cite{Heemskerk:2009pn,Penedones:2010ue,Paulos:2011ie,Fitzpatrick:2010zm,Fitzpatrick:2011ia,Fitzpatrick:2011hu,Fitzpatrick:2011dm,Paulos:2016fap,Komatsu:2020sag,Li:2021snj,Gadde:2022ghy,vanRees:2022zmr,vanRees:2023fcf,Marotta:2024sce,Bagchi:2023fbj,Alday:2024yyj,Lipstein:2025jfj}. However, it seems closely related to the prescriptions put forward in \cite{Komatsu:2020sag} (see also \cite{Berenstein:2025tts}), although our version keeps boundary insertions \textit{real} at every step. It will be interesting to compare the results of these different approaches in detail. The computation leading to \eqref{limit inner product} provides the simplest example of this limiting procedure, as it indeed corresponds to the 1-to-1 scattering process. We note that the precise implementation of step 3 above requires appropriate rescaling by powers of $\ell\sim c^{-1}$, and the limiting procedure is to be understood in the sense of distributions.\footnote{Similar manipulations are performed in \cite{Nguyen:2023miw} in a slightly different context.}
	
	The present work left aside the contraction of lowest-weight UIRs to \textit{massless} Poincaré UIRs. This will be the subject of the second paper in this series \cite{Nguyen:appear}. While Evans established that the way massive particles arise from lowest-weight representations is essentially unique \cite{Evans:1967zgs}, this is not obviously the case for the contraction to massless Poincaré UIRs \cite{Evans:1967zgs,Angelopoulos:1980wg}. One possible way forward was recently suggested in \cite{Berenstein:2025qhb,Berenstein:2025tts}: a double scaling limit in which an infinite boost $\eta \to \infty$ is implemented along $\ell \to \infty$, with fixed boosted energy\footnote{Comparison of \eqref{flat momentum param} and \eqref{momentum rapidity param} shows that $\cosh \eta=\frac{1+x^2}{1-x^2}$, so that infinite boost $\eta \to \infty$ corresponds to insertion on the unit sphere $x \to 1$ in $\mathbb{R}^3$.} 
	\begin{equation}
		\label{double scaling}
		p^0=\frac{\Delta}{\ell} \cosh \eta\,, \qquad (\eta, \ell \to \infty)\,.
	\end{equation}
	In a recent paper, we described the contraction of massive particle UIRs to massless ones in the infinite boost limit $\eta \to \infty$ \cite{Nguyen:2026yho}, independently of any flat limit contraction $\mathfrak{so}(2,3) \to \mathfrak{iso}(1,3)$. The result is simple:
	the $(2s+1)$ spin states of a massive spin-$s$ UIR converge to the direct sum of massless particle UIRs with helicities ranging from $-s$ to $s$ in integer steps,
	\begin{equation}
		\label{direct sum bis}
		\operatorname{V}_{m,s} \quad \stackrel{\eta \to \infty}{\longrightarrow} \quad \bigoplus_{J=-s}^s \operatorname{V}_{0,J}\,.
	\end{equation}
	The double-scaling limit \eqref{double scaling} can be implemented in two consecutive steps: first the contraction of $\mathcal{D}(\Delta,s)$ to a massive Poincaré UIR as described in the present paper, followed by the infinite boost limit yielding \eqref{direct sum bis}. In this case, we must allow arbitrarily large $\Delta \sim m \ell \to \infty$ in the first part of the process. However, the double scaling limit \eqref{double scaling} may also be implemented at fixed $\Delta$ by correlating the two limits through $\ell \sim e^\eta$, which is what is actually suggested in \cite{Berenstein:2025qhb,Berenstein:2025tts}. This approach is especially important given that the spectrum of scaling dimensions in concrete examples of AdS/CFT dualities cannot be tuned at will \cite{Aharony:1999ti}. Even more importantly, these concrete examples always feature \textit{short multiplets} dual to linearized gauge fields of the corresponding AdS supergravities, whose treatment is not accommodated by the present analysis. The complete treatment of the flat limit contraction of $\mathcal{D}(\Delta,s)$ to massless particle representations, at fixed $\Delta$ and including the special case of short multiplets ($\Delta=s+1$), will be presented in \cite{Nguyen:appear}.
	
	As a final comment, we believe that developing a first-principles understanding of the flat limit of AdS/CFT, rooted in the Hilbert space contraction described here, will allow progress regarding the consistent formulation of \textit{flat-space holography}, in its celestial \cite{Pasterski:2021raf,McLoughlin:2022ljp,Donnay:2023mrd} and/or carrollian incarnations \cite{Bagchi:2025vri,Nguyen:2023vfz,Nguyen:2025sqk,Nguyen:2025zhg,Ruzziconi:2026bix}.

	\section*{Acknowledgments}
	
	We thank Slava Rychkov and Joan Simon for stimulating discussions. This work is supported by a postdoctoral research fellowship of the
	F.R.S.-FNRS (Belgium).
	
	\appendix
	
	\section{Inner product of pseudo-momentum states}
	\label{app inner product}
	Here, we give the explicit derivation of \eqref{pseudo-momentum kernel}. By the definition \eqref{pseudo momentum state}, we have
	\begin{equation}
		\label{kernel appendix}
		\langle p_1,sh_1|p_2,sh_2\rangle=\langle \Delta,sh_1|U(L^{-1}(\hat p_1))U(L(\hat p_2)) |\Delta,sh_2\rangle=\langle \Delta,sh_1|U(H) |\Delta,sh_2\rangle\,,
	\end{equation}
	where we defined the group element $H\equiv L^{-1}(\hat p_1) L(\hat p_2) \in \operatorname{SO}(1,3)$. We may decompose it into a pure boost and a rotation,
	\begin{equation}
		\label{H definition}
		H=L(\hat q_{12})W_{12}\,, \qquad W_{12}\equiv L^{-1}(\hat q_{12})H\,,
	\end{equation}
	where $\hat q_{12}$ is defined as 
	\begin{equation}
		\label{q12 definition}
		\hat q_{12}\equiv H \hat k=L^{-1}(\hat p_1) \hat p_2\,.    
	\end{equation}
	We can check that $W_{12}$ is indeed a pure rotation, i.e., that it leaves the reference momentum $\hat k^\mu=(1,\vec 0)$ invariant,
	\begin{equation}
		W_{12}\, \hat k=L^{-1}(\hat{q}_{12})H\hat k=L^{-1}(\hat{q}_{12})\hat q_{12}=\hat k\,.
	\end{equation}
	Because it is a pure rotation, its action of the primary states is known and allows to rewrite \eqref{kernel appendix} as
	\begin{equation}
		\label{kernel appendix 2}
		\langle p_1,sh_1|p_2,sh_2\rangle=\sum_{h'} [W_{12}]_{h'h_2}\, \langle \Delta,sh_1|U(L(\hat q_{12})) |\Delta,sh'\rangle  \,.
	\end{equation}
	Now let us parametrize the unit vector $\hat q_{12}$ as 
	\begin{equation}
		\hat q_{12}^\mu=(\cosh \eta\,,\sinh \eta\, \vec n)\,, \qquad |\vec n|=1\,,
	\end{equation}
	such that $L(\hat q_{12})$ is a boost along $\vec n$,
	\begin{equation}
		\label{boost along n}
		L(\hat q_{12})=\exp \left(i \eta\,  n^i L_{0i}\right)\,.
	\end{equation}
	To compute the matrix element of $\langle \Delta,sh_1|U(L(\hat q_{12})) |\Delta,sh'\rangle$, it is enough to consider the $\mathfrak{sl}(2)$ subalgebra generated by
	\begin{equation}
		K^\pm\equiv n^i L_i^\pm\,, \qquad D\,,
	\end{equation}
	which satisfy the commutation relations
	\begin{equation}
		[D\,,K^\pm]=\pm K^\pm\,, \qquad [K^-,K^+]=2 D\,.
	\end{equation}
	The boost element \eqref{boost along n} obeys the disentangling identity \cite{Perelomov:1986tf}
	\begin{equation}
		\begin{split}
			L(\hat q_{12})&=\exp \left(\tfrac{\eta}{2} (K^+-K^-) \right)\\
			&=\exp \left(\tanh \tfrac{\eta}{2}\, K^+\right) \exp\left(-2 \ln \cosh \tfrac{\eta}{2}\, D\right) \exp\left(-\tanh \tfrac{\eta}{2}\, K^-\right)\,,
		\end{split}
	\end{equation}
	such that, recalling $K^+=(K^-)^\dagger$ and using the lowest-weight condition \eqref{primary condition} on the bra and the ket, we obtain
	\begin{equation}
		\label{kernel appendix 3}
		\begin{split}
			\langle \Delta,sh_1|U(L(\hat q_{12})) |\Delta,sh'\rangle&=\langle \Delta,sh_1|U(\exp\left(-2 \ln \cosh \tfrac{\eta}{2}\, D\right)) |\Delta,sh'\rangle\\
			&=\left(\cosh \tfrac{\eta}{2}\right)^{-2\Delta} \langle \Delta,sh_1|\Delta,sh'\rangle=\left( \frac{2}{1+\cosh \eta}\right)^\Delta\, \delta_{h_1h'}\,.
		\end{split}
	\end{equation}
	As a last step, we write
	\begin{equation}
		\label{cosh eta appendix}
		\cosh \eta=\hat q_{12}^0=-\hat p_1 \cdot \hat p_2\,,
	\end{equation}
	where the last equality is derived below in \eqref{q12 components}. Putting together \eqref{kernel appendix 2}, \eqref{kernel appendix 3}, and \eqref{cosh eta appendix}, yields
	\begin{equation}
		\langle p_1,sh_1|p_2,sh_2\rangle=\left( \frac{2}{1-\hat p_1 \cdot \hat p_2}\right)^\Delta\, [W_{12}]_{h_1h_2}\,,
	\end{equation}
	which is the identity we intended to prove.
	
	\section{Wigner rotation as a product of inversions}
	\label{app: Wigner rotation}
	
	Here we prove the assertion
	\begin{equation}
		\label{II=W}
		\mathcal{I}\indices{_i^k}(\vec x)\, \mathcal{I}
		_{kj}(\vec x/x^2-\vec y)=[W_{12}]_{ij}\,,
	\end{equation}
	where
	\begin{equation}
		\mathcal{I}_{ij}(\vec x)=\delta_{ij}-\frac{2 x_i x_j}{x^2}\,,
	\end{equation}
	and $[W_{12}]_{ij}$ is the vector representation of the Wigner rotation 
	\begin{equation}
		\label{W12}
		W_{12}=L^{-1}(\hat{q}_{12})L^{-1}(\hat p_1)L(\hat p_2)\,, \qquad \hat q_{12}=L^{-1}(\hat p_1) \hat p_2\,. 
	\end{equation}
	
	\paragraph{Left-hand side.} 
	First, explicitly evaluating the left-hand side of \eqref{II=W} directly yields
	\begin{equation}
		\label{S A decomposition}
		\mathcal{I}\indices{_i^k}(\vec x)\, \mathcal{I}
		_{kj}(\vec x/x^2-\vec y)=\delta_{ij}+S_{ij}+A_{ij}\,,
	\end{equation}
	where
	\begin{equation}
		\begin{split}
			S_{ij}&=-\frac{2}{D}\left(x_i x_j y^2+y_i y_j x^2-\vec x \cdot \vec y(x_i y_j+x_j y_i)\right)\,,\\
			A_{ij}&=-\frac{2}{D}\, (1-\vec x \cdot \vec y)(x_i y_j-x_j y_i)\,,
		\end{split}
	\end{equation}
	and
	\begin{equation}
		\label{denominator}
		D=1-2 \vec x \cdot \vec y+x^2 y^2\,.
	\end{equation}
	
	\paragraph{Right-hand side.}
	Explicit evaluation of the right-hand side of \eqref{II=W} is more challenging. First, we establish some useful properties of the Lorentz group element $H\equiv L^{-1}(\hat p_1)L(\hat p_2)$.
	On the one hand, it satisfies
	\begin{equation}
		H\indices{^\mu_0}=\hat q_{12}^\mu\,,
	\end{equation}
	which directly follows from \eqref{q12 definition}.
	Since it is an element of the Lorentz group, it also satisfies
	\begin{equation}
		0=\eta_{0i}=H\indices{^\mu_0} \eta_{\mu\nu} H\indices{^\nu_i}=-H\indices{^0_0}H\indices{^0_i}+H\indices{^j_0}H_{ji}=-\hat q_{12}^0 H\indices{^0_i}+\hat q_{12}^j H_{ji}\,,
	\end{equation}
	or equivalently
	\begin{equation}
		\hat q_{12}^j H_{ji}=\hat q_{12}^0 H\indices{^0_i}\,,
	\end{equation}
	which will greatly simplify the following computations.
	
	We now proceed to the explicit evaluation of the spatial components $[W_{12}]_{ij}$ using the explicit form of the standard boost matrix $L(\hat p)\indices{^\mu_\nu}$ \cite{Weinberg:1995mt,Iacobacci:2024laa}
	\begin{equation}
		\label{boost matrix}
		\begin{split}
			L(\hat p)\indices{^0_0}&=\hat p^0\,,\\
			L(\hat p)\indices{^i_0}&=L(\hat p)\indices{^0_i}=\hat p^i\,,\\
			L(\hat p)\indices{^i_j}&=\delta^i_j+(\hat p^0+1)^{-1}\, \hat p^i \hat p_j\,,
		\end{split}
	\end{equation}
	with inverse
	\begin{equation}
		\begin{split}
			L^{-1}(\hat p)\indices{^0_0}&=\hat p^0\,,\\
			L^{-1}(\hat p)\indices{^i_0}&=L^{-1}(\hat p)\indices{^0_i}=-\hat p^i\,,\\
			L^{-1}(\hat p)\indices{^i_j}&=\delta^i_j+(\hat p^0+1)^{-1}\, \hat p^i \hat p_j\,.
		\end{split}
	\end{equation} 
	Using these, we compute
	\begin{equation}
		\label{q12 components}
		\begin{split}
			\hat q_{12}^0&=-\hat p_1 \cdot \hat p_2\,,\\
			\hat q_{12}^i&=\hat p_2^i-\hat p_1^i\left(\hat p_2^0-(\hat p_1^0+1)^{-1} \hat p_1^j \hat p_{2j} \right)\,,
		\end{split}
	\end{equation}
	and
	\begin{equation}
		\begin{split}
			H\indices{^0_i}&\equiv [L^{-1}(\hat p_1)L(\hat p_2)]\indices{^0_i}=-\hat p_{1i}+\hat p_{2i}\left(\hat p_1^0-(\hat p_2^0+1)^{-1} \hat p_1^j \hat p_{2j} \right)\,,\\
			H\indices{^i_j}&\equiv [L^{-1}(\hat p_1)L(\hat p_2)]\indices{^i_j}=\delta^i_j+\frac{\hat p_1^i \hat p_{1j}}{\hat p_1^0+1}+\frac{\hat p_2^i \hat p_{2j}}{\hat p_2^0+1} +\left(\frac{\hat p_1^k  \hat p_{2k}}{(\hat p_1^0+1)(\hat p_2^0+1)}-1\right) \hat p_1^i \hat p_{2j}\,.
		\end{split}
	\end{equation}
	Putting everything together, we evaluate the spatial components of \eqref{W12},
	\begin{equation}
		\begin{split}
			[W_{12}]\indices{^i_j}&=-\hat q_{12}^i H\indices{^0_j}+(\delta^{ik}+\left(\hat q_{12}^0+1)^{-1}\, \hat q_{12}^i \hat q_{12}^k\right)H_{kj}=H\indices{^i_j}-\frac{\hat q_{12}^i}{\hat q_{12}^0+1} H\indices{^0_j}\\
			&=\delta\indices{^i_j}+\bar S\indices{^i_j}+\bar A\indices{^i_j}\,,
		\end{split}
	\end{equation}
	where
	\begin{equation}
		\begin{split}
			\bar S_{ij}&=-\frac{2}{\Sigma}\left[\hat p_{1i} \hat p_{1j}\, (\hat p_{2k} \hat p_2^k)+\hat p_{2i} \hat p_{2j}\, (\hat p_{1k} \hat p_1^k)-(\hat p_{1i} \hat p_{2j}+\hat p_{2i} \hat p_{1j})\hat p_1^k \hat p_{2k}  \right]\,,\\
			\bar A_{ij}&=-\frac{2}{\Sigma}\left[(\hat p_1^0+1)(\hat p_2^0+1)-\hat p_1^k \hat p_{2k} \right]\left(\hat p_{1i} \hat p_{2j}-\hat p_{2i} \hat p_{1j} \right)\,,
		\end{split}
	\end{equation}
	such that
	\begin{equation}
		\Sigma = 2 (\hat p_1^0+1)(\hat p_2^0+1)(1-\hat p_1 \cdot \hat p_2)\,.
	\end{equation}
	Finally plugging in $\hat p_1=\hat p(\vec x)$ and $\hat p_2=\hat p(\vec y)$ as parameterized as in \eqref{flat momentum param}, and noticing that $\hat p^i(\vec x)=(\hat p^0(\vec x)+1)x^i$, we easily establish
	\begin{equation}
		\Sigma=(\hat p_1^0+1)^2(\hat p_2^0+1)^2 D\,,
	\end{equation}
	and
	\begin{equation}
		\bar S_{ij}=S_{ij}\,, \qquad \bar A_{ij}=A_{ij}\,.
	\end{equation}
	This proves the assertion \eqref{II=W}.
	
	\paragraph{Generalization to rank-$s$ tensor representations.} The equality \eqref{II=W} generalizes to
	\begin{equation}
		\mathcal{I}\indices{_I^K}(\vec x)\, \mathcal{I}
		_{KJ}(\vec x/x^2-\vec y)=[W_{12}]_{IJ}\,,
	\end{equation}
	with multi-indices $I=i_1...\,i_s$.
	Indeed, it is obtained as the totally symmetric and traceless part of the $s$-fold product of \eqref{II=W}. 
	
	\paragraph{Rotation axis and angle.} The expression \eqref{S A decomposition} may be compared with Rodrigues' formula for the rotation by an angle $\theta$ around the unit vector $\vec n$,
	\begin{equation}
		\label{Rodrigues}
		R_{ij}(\theta,\vec n)=\delta_{ij}+(1-\cos \theta)(n_i n_j-\delta_{ij})-\sin \theta\, \epsilon_{ijk} n^k\,.
	\end{equation}
	Equating the trace of \eqref{S A decomposition} and \eqref{Rodrigues} determines the angle $\theta$ through
	\begin{equation}
		\cos \theta=1+\frac{\delta^{ij}S_{ij}}{2}=1-\frac{2(x^2 y^2-(\vec x \cdot \vec y)^2)}{D}=1-\frac{2|\vec x \times \vec y|^2}{D}\,,
	\end{equation}
	written in terms of the cross product $(\vec x \times \vec y)_i=\varepsilon_{ijk} x^j y^k$. Equating the antisymmetric parts yields
	\begin{equation}
		\sin \theta\, n_k=-\frac{1}{2} \varepsilon_{kij} A^{ij}=\frac{2(1-\vec x \cdot \vec y)(\vec x \times \vec y)_k}{D}\,,
	\end{equation}
	or equivalently
	\begin{equation}
		\vec n=\frac{\vec x \times \vec y}{|\vec x \times \vec y|}\,, \qquad \sin \theta=\frac{2(1-\vec x \cdot \vec y)|\vec x \times \vec y|}{D}\,.
	\end{equation}
	At this point, one can perform a consistency check and verify that $\cos^2 \theta+ \sin^2 \theta=1$, which is indeed the case by virtue of the identity $D=|\vec x \times \vec y|^2+(1-\vec x \cdot \vec y)^2$. Using the half-angle formulas, we may also write 
	\begin{equation}
		\cos \frac{\theta}{2}=\frac{1-\vec x \cdot \vec y}{\sqrt{D}}\,, \qquad \sin \frac{\theta}{2}=\frac{|\vec x \times \vec y|}{\sqrt{D}}\,, \qquad \tan \frac{\theta}{2}=\frac{|\vec x \times \vec y|}{1-\vec x \cdot \vec y}\,.
	\end{equation}
	In summary, the matrix given in \eqref{S A decomposition} is indeed that of a pure rotation $R(\theta,\vec n)$, with rotation angle and rotation axis which we just determined. Note that the rotation axis $\vec x \times \vec y$ is also the axis $\vec p_1 \times \vec p_2$.
	
	\bibliography{bibl}
	\bibliographystyle{JHEP}  
\end{document}